\documentstyle[12pt,amssymb,epsfig,citesort]{article}
\begin{document}

\def\alps{\alpha_s}
\def\as{\frac{\alpha_s}{4 \pi}}
\def\msbar{${\rm{\overline{MS}}}$}
\def\alps1{${\cal{O}}(\alpha_s^1)$}
\def\alp{${\cal{O}}(\alpha_s^0)$}
\def\alpsq{${\cal{O}}(\alpha_s^2)$}
\def\Qt{{\tilde{Q}}}
\def\oalps{${\cal{O}}(\alpha_s)$}

\setlength{\parskip}{0.45cm}
\setlength{\baselineskip}{0.75cm}
%
%
\begin{titlepage}
\begin{flushright}
DO-TH 99/18  \\
September 1999 \\
\end{flushright}
\begin{center}
\vspace*{1.8cm}
{\LARGE
\hbox to\textwidth{\hss
{\bf Massless Parton Asymptotics within} \hss}

\vspace{+0.2cm}
\hbox to\textwidth{\hss
{\bf Variable Flavour Number Schemes} \hss}}

\vspace*{2.0cm} 
{\large S.\ Kretzer}\\
{Institut f\"{u}r Physik, Universit\"{a}t Dortmund,
D-44221 Dortmund, Germany}\\ 
\vspace*{3.6cm}
{\bf Abstract}

\vspace{-0.3cm}
\end{center}
%
In this note we formulate and investigate theoretical uncertainties
for high $Q^2$ deep inelastic heavy quark
(charm, etc.) production rates 
which arise within collinear resummation techniques from variations of the 
{\it a priori} unknown charm input scale $Q_0$
of \oalps\ variable flavour number schemes. 
We show that $Q_0$ variations constitute a source of considerable theoretical 
uncertainty of present \oalps\ calculations within such schemes
and we suggest to consider a scale optimization from higher order corrections. 
We also outline how 
the stability of the fixed order and collinearly resummed perturbation series for
heavy quark production can be comparatively investigated by variation of $Q_0$.

\end{titlepage} 
The present discussion on the appropriate scheme for the perturbative treatment
of the deep inelastic production of heavy quarks of mass $m\gg \Lambda_{QCD}$
can be partly
traced back to the question what is the effective
expansion parameter for high $Q^2$ predictions. While fixed order perturbation theory
(FOPT) proceeds strictly stepwise in powers of 
$(\alpha_s / 2\pi )$ at all scales, variable flavor number schemes
(VFNSs) are based upon the expectation that terms  
$\sim (\alpha_s / 2\pi \times\ln Q^2/m^2)^n$ from collinear regions in the phase 
space have to be resummed \cite{svz} to all perturbative orders $n$ for high $Q^2$ when 
$(\alpha_s / 2\pi \times\ln Q^2/m^2) \rightarrow {\cal{O}}(1)$. Such terms are 
undebatedly 
present in the high $Q^2$ limit of the perturbative partonic coefficient functions 
but their impact is less clear \cite{grs}  
on observable hadronic quantities like the charm component of
the deep inelastic structure function $F_2^c$ where
the partonic coefficient functions have to be convoluted with modern, i.e.\ steep,
parton distribution functions $\stackrel{(-)}{q}(x,\mu^2_F)$ and
dominantly $g(x,\mu^2_F)$, $\mu_f\sim m$.
The question which ordering of the perturbation series optimizes its convergence
can therefore not be answered {\it a priori} but only from explicit 
quantitative, i.e.\ numerical
investigations \cite{grs,or}; prominent tools for testing  
perturbative stability being $K$ factor considerations \cite{grs} or 
scale variations \cite{or}. At present both criteria indicate a well
behaved fixed order perturbation series for
relevant subasymptotic but large scales $Q \gg m$ 
\cite{grs,or,grv98,gkr3}. As regards scale uncertainties, 
mainly variations of the mass factorization scale $\mu_F$
have been considered so far despite the fact 
that collinear resummation techniques introduce an additional 
arbitrary scale in the process set by the input scale $Q_0$
for the heavy quark:      

\noindent
Recently proposed variable flavour number schemes 
\cite{aot,acot,collins,mrrs,thoro}
for global PDF analyses
are constructed upon the boundary condition
\begin{equation}
\label{boundary}
\left. q_{{}\atop{\tiny{H}}}^{(n_f+1)}(x,Q_0^2)\right|_{Q_0=m} = 0 
\end{equation}
for a heavy sea quark density to enter the massless partonic 
renormalization group (RG)-evolution equations
which resum collinear splitting subdiagrams to all orders at the price of neglecting
mass dependent terms. 
In Eq.\ (\ref{boundary}) $m$ is the heavy quark mass and
the heavy quark input scale $Q_0$ is in more technical terms the transition
(or {\it switching}) scale from a factorization scheme with $n_f$ to the one
with $n_f +1$ partonic quark degrees of freedom \cite{ct}. 
Since the scale $Q_0$ is of no physical meaning, a RG-like equation
\begin{equation}
\label{rgelike} 
\frac{\partial\ {\cal O}}{\partial \ln Q_0^2} = 0
\end{equation} 
holds {\it ideally} for any heavy quark observable ${\cal O}$. At limited 
perturbative order, Eq.\ (\ref{rgelike}) will obviously
be violated to some extent 
which we will investigate below for the charm contribution to the NC structure
function ${\cal O}=F_2^c$.   

At the heart of the variable flavour number schemes of 
\cite{aot,acot,collins,mrrs,thoro}
is some interpolation prescription between fixed order perturbation 
theory, assumed to be valid around $Q^2= {\cal{O}}(m^2)$,
and the $Q^2\gg m^2$ massless parton (MP) 
asymptotics derived from the boundary condition
in Eq.\ (\ref{boundary}). To avoid within our rather general 
considerations a discussion of the peculiarities
of the distinct 
heavy quark schemes we denote such interpolations very 
schematically as
\begin{equation}
\label{simple}
{\rm VFNS} = w(m^2/Q^2)\times {\rm FOPT} + [1-w(m^2/Q^2)]\times {\rm MP}; 
\ \ \ w\rightarrow
\left\{
\begin{array}{l} 1,\ m^2/Q^2\rightarrow 1 \\ 0,\ m^2/Q^2 \rightarrow 0
\end{array}\right. 
\end{equation}
where the simple weight $w$ is meant to represent all the details of 
some elaborate scheme prescription \cite{aot,acot,collins,mrrs,thoro,bmsn}. 
The deviation of VFNS from FOPT is thus {\it normalized} to MP and 
the predictive power of VFNS in Eq.\ (\ref{simple}) depends on the stability 
of the asymptotic MP prediction which is obtained from
the boundary (\ref{boundary}) at $Q_0=m$ via massless RG evolutions.
Equation (\ref{boundary}) emerges from the matching conditions
of a factorization scheme with $n_f$ active flavours
to a scheme with $n_f+1$ active flavours at some {\it{a priori}} arbitrary
transition (or {\it{switching}}) scale $Q_0$. The general transformation
equations for quark ($q$) and gluon ($g$) parton densities as well as for
$\alpha_s$ read up to NLO \cite{ct,bmsn}
\begin{equation}
\label{bound}
\begin{array}{lllll} 
q_{{}\atop{\tiny{H}}}^{(n_f+1)}(x,Q_0^2) &=&
\frac{\alpha_s(Q_0^2)}{2 \pi}
\ \ln \frac{Q_0^2}{m^2} 
\ \int_{x}^1\frac{d\xi}{\xi}\ P_{qg}^{(0)}(\xi)   
\ g^{(n_f)}\left(\frac{x}{\xi},Q_0^2\right)
&+& {\cal{O}}(\alpha_s^2) \\ 
g^{(n_f+1)}(x,Q_0^2) &=&  g^{(n_f)}(x,Q_0^2)\ \left(
1+\frac{\alpha_s(Q_0^2)}{6\pi}\ \ln
\frac{m^2}{Q_0^2}\right) 
&+& {\cal{O}}(\alpha_s^2) \\ 
\alpha_s^{(n_f+1)}(Q_0^2) &=& \alpha_s^{(n_f)}(Q_0^2)\ \left/
\ \left( 1 + \frac{\alpha_s(Q_0^2)}{6\pi} \ln
\frac{m^2}{Q_0^2}\right) \right.
&+& {\cal{O}}(\alpha_s^3) \\ 
q^{(n_f+1)}(x,Q_0^2) &=& q^{(n_f)}(x,Q_0^2) &+& {\cal{O}}(\alpha_s^2) 
\end{array}
\end{equation}
and obviously reduce to Eq.\ (\ref{boundary}) for $Q_0=m$:
\begin{equation} 
\label{boundatm}
q_{{}\atop{\tiny{H}}}^{(n_f+1)}(x,m^2) = 0 \ ,
\ g^{(n_f+1)}(x,m^2) =  g^{(n_f)}(x,m^2) \ , 
\ \alpha_s^{(n_f+1)}(m^2) = \alpha_s^{(n_f)}(m^2)\ .
\end{equation}
In Eqs.\ (\ref{bound}) and (\ref{boundatm}) $q_{{}\atop{\tiny{H}}}$ introduces  
a partonic heavy quark density into the massless evolution equations and
the unspecified $\alpha_s(Q_0^2)$ may be either $\alpha_s^{(n_f)}(Q_0^2)$ or 
$\alpha_s^{(n_f+1)}(Q_0^2)$ because the difference is of the orders neglected
in (\ref{bound}). 
The argument of a continuous $g(x,Q_0^2)$ and $\alpha_s(Q_0^2)$ 
has been advanced \cite{ct,acot} 
for adopting $Q_0=m$ in all NLO parton distribution sets constructed
so far along a VFNS philosophy \cite{grv92,ref:mrst,cteq5}. On
the other hand, when $Q_0 \neq m$
the discontinuity of $\alpha_s$ is in practice as small as
maximally 4\% up to $Q_0^2$ as high as 1000 ${\rm GeV}^2$.
Anyway, the recently completed NNLO transformation equations \cite{bmsn,alpsnnlo} 
reveal that the possibility of a continuous evolution 
across $Q_0$ breaks down beyond NLO by nonlogarithmic higher order corrections
to (\ref{bound}), (\ref{boundatm}). 
The restriction to $Q_0=m$ should hence be abandoned and
effects of 
varying $Q_0$
should be taken into account on the same
level as the variations of the mass factorization scale $\mu^2_F$
usually considered. 
Indeed, along
\begin{equation}
\ln \frac{Q^2}{m^2}\ =\ \ln \frac{Q_0^2}{m^2}\ +\ \ln \frac{\mu^2_F}{Q_0^2}
\ +\ \ln \frac{Q^2}{\mu^2_F}
\end{equation}
the two scales $Q_0$ and $\mu_F$
define quite symmetrically which portion of the quasi-collinear 
$\ln (Q^2/m^2)$ is actually
resummed [$\ln (\mu^2_F/Q_0^2)$] 
and what amount is kept at fixed order,
either in the boundary condition for $q_{{}\atop{\tiny{H}}}$ 
[$\ln (Q_0^2/m^2)$] or in the 
hard scattering coefficient function $C_2^g$ in Eq.\ (\ref{mp})
below [$\ln (Q^2/\mu^2_F)$].  
We will investigate the residual $Q_0$ dependence   
for the charm production contribution to the deep
inelastic structure function $F_2$ using $m=m_c(=1.5\ {\rm{GeV}})$. To
avoid complications from an interplay of several scales we
will decouple the bottom and top quark from the process ($m_{b,t}\rightarrow \infty$)
and we will fix the factorization scale at $\mu_F = Q$.  

In the asymptotic limit $m_c^2/Q^2\rightarrow 0$ the 
schemes \cite{acot,mrrs,thoro} reduce - as in Eq.\ (\ref{simple}) -
to the so-called
{\it Zero Mass Variable Flavour Number Scheme}, equivalent to the `massless parton' 
scenario of Ref.\ \cite{grs} where any mass dependence is dropped except
for the boundary conditions in (\ref{bound}). 
We will consider such a scenario in the
following and ignore terms of ${\cal{O}}(m_c^2/Q^2)$ because these are not 
handled uniformly in the individual realizations \cite{acot,mrrs,thoro} of a 
VFNS.\footnote{Indeed the charm scheme of \cite{thoro} is explicitly constructed
upon $Q_0=m$ and a generalization of this particular 
scheme for $Q_0\neq m$ seems nontrivial.} 
For definiteness we consider an $F_2^{c,MP}$ in $\gamma^\ast P$
scattering which is given by
\begin{equation}
\label{mp}
\frac{1}{x} F_2^{c,MP}(x,Q^2) 
= e_c^2 \left\{ \left(c+{\bar{c}} \right)(x,Q^2)
+ \frac{\alpha_s(Q^2)}{2\pi} \left[ g\otimes C_2^g + \left( c+{\bar{c}} \right)
\otimes C_2^q\right](x,Q^2)\right\}\ \ \ ,
\end{equation} 
where the $C_2^{g,q}$ are the massless \msbar\ coefficient functions  
\cite{aem,fupe}. It has already been pointed out in \cite{grs} that the `massless
parton' $F_2^{c,MP}$ in Eq.\ (\ref{mp}) can be rather arbitrarily suppressed if some 
larger {\it{effective}} charm mass is introduced \cite{mrs93} into the boundary 
condition (\ref{boundatm}). 
Our investigation here will clarify the situation 
if - for a fixed value of the {\it{physical}} charm mass $m_c$ 
- the {\it{unphysical}} switching scale $Q_0$ is varied 
consistently according to the NLO boundary equations (\ref{bound}).
\begin{figure}[t]
\vspace*{-0.5cm}
\hspace*{-1.25cm}
\epsfig{figure=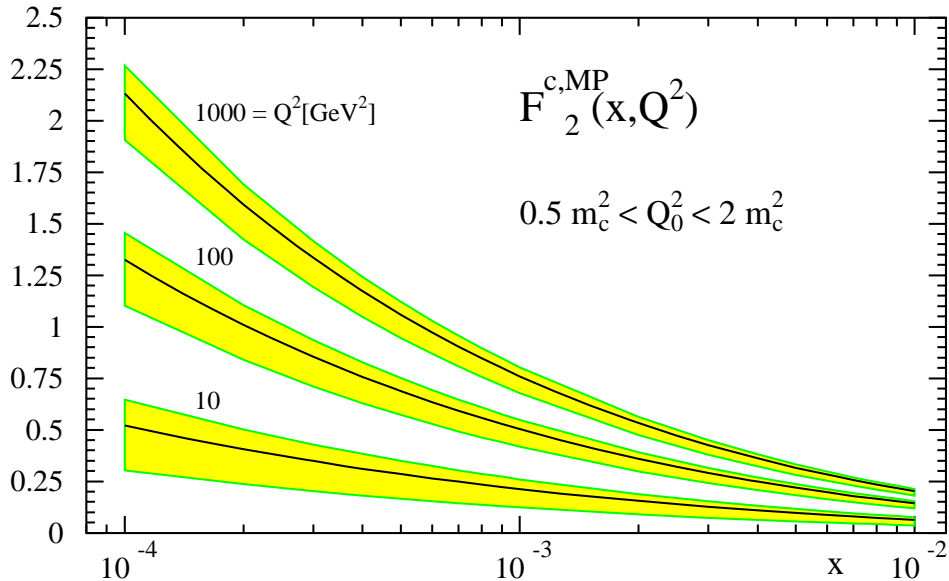,width=16cm}
\vspace*{-2.5cm}
\caption{\label{switch}
Dependence of the `massless parton'-type charm structure function
$F_2^{c,MP}$ in Eq.\ (\ref{mp}) on the switching scale $Q_0$ in the boundary 
conditions in Eq.\ (\ref{bound}). The solid lines represent the
central value of $Q_0=m_c$ and $F_2^{c,MP}$ decreases monotonically with
increasing $Q_0$. The underlying parton distributions below $Q_0$
are those of Ref.\ \cite{grv92}.}
\end{figure} 
Fig.\ \ref{switch} shows the effect if the transition scale is allowed to
vary over the range $m_c^2/2<Q_0^2<2\ m_c^2$ where the central value $Q_0=m_c$ is
represented by the solid lines and $F_2^{c,MP}$ monotonically decreases with 
increasing $Q_0$.\footnote{
Allowing for $Q_0<m_c$ in Eq.\ (\ref{bound}) leads obviously to $c(x,Q_0^2)<0$
which appears
somewhat counter-intuitive in probabilistic parton model language.
Note, however, that a negative charm input arises 
even for $Q_0=m$ from higher
order corrections to Eq.\ (\ref{bound}) \cite{smith}. Anyway, the 
{\it measurable} cross section $F_2^c$ is certainly positive above the 
physical threshold $Q^2(1/x-1)>4m_c^2$.}
The evolution leading to the results in Fig.\ \ref{switch} is based
on the $n_f=3$ valence-like 
NLO input of Ref.\ \cite{grv92} using NLO (2-loop) splitting functions. 
Above $Q_0$ the 
evolution deviates from \cite{grv92} because we consider general
$Q_0 \neq m_c$ here and we ignore - as mentioned above - 
any bottom quark effects ($m_b\rightarrow \infty$). 
The amount of change of $F_2^{c,MP}$ under variation of $Q_0$
hints at a reasonable perturbative stability. Nevertheless, 
the error represented by the shaded bands in Fig.\ \ref{switch}
is of the typical order of discrepancies between VFNS and
FOPT calculations \cite{bmsn,thoro} which questions the gain in
predictivity if FOPT is abandoned for VFNS.  
Such uncertainties are critical for precise charm predictions to compare
with future experimental accuracy, 
especially at experimentally most relevant
intermediate scales and regarding the fact that $Q_0^2$ was only allowed 
to fluctuate by a factor of $2$. This latter limitation rests on the 
assumption 
that $Q_0$ has to be very close to $m_c$ for the all-order logarithms
$(\alpha_s/2\pi \times \ln Q^2/m_c^2)^n$ to be correctly resummed. One can 
as well adopt a very different point of view towards the 
choice of $Q_0$. One can easily see that inserting Eq.\ (\ref{bound}) into
Eq.\ (\ref{mp}) gives
\begin{equation}
\label{foptlim}
\lim_{Q_0\rightarrow Q} F_2^{c,MP}(x,Q^2)=\left[
F_2^{GF}\otimes\left(
1+\frac{\alpha_s}{2\pi} C_2^q\right)\right](x,Q^2)
+{\cal{O}}\left(\alpha_s^2\right)
+{\cal{O}}\left(\frac{m_c^2}{Q^2}\right)
\end{equation}
which is dominated by the ${\cal O}[\alpha_s \ln (Q^2/m_c^2)]$
LO gluon fusion term $F_2^{GF}$ of the fixed order 
perturbation series. We may
hence consider $Q_0\rightarrow Q$ in a sense
as a continuous path from variable flavour number to fixed order calculations.
We should then consider values of $Q_0$ as high as we trust fixed order 
perturbation theory. 
\begin{figure}[t]
\vspace*{-0.5cm}
\hspace*{-1.25cm}
\epsfig{figure=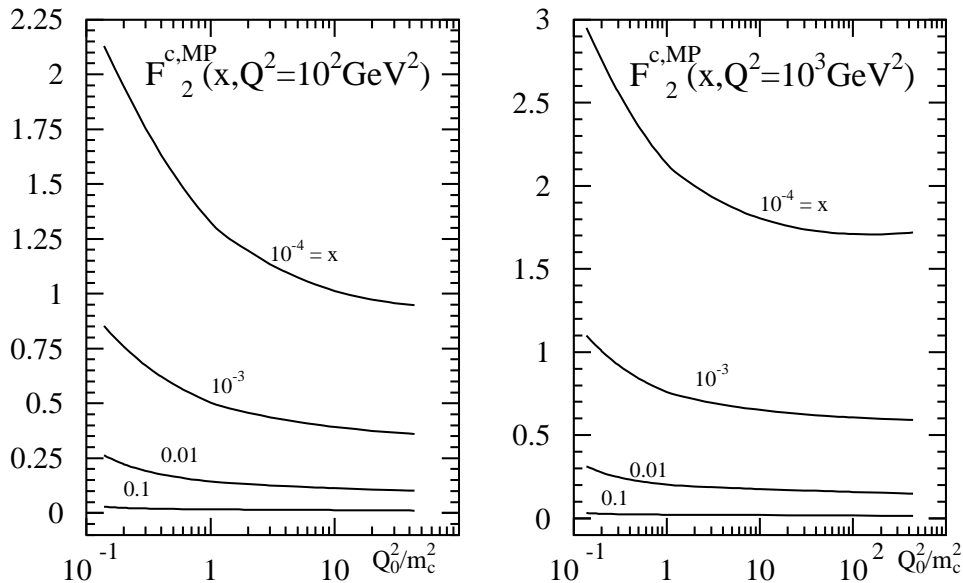,width=16cm}
\vspace*{-2.5cm}
\caption{\label{switch2}
Dependence of the `massless parton'-type charm structure function
$F_2^{c,MP}$ in Eq.\ (\ref{mp}) on the switching scale $Q_0$ in the boundary 
conditions in Eq.\ (\ref{bound}). In this Figure $Q_0$ is allowed to vary
(maximally)
between the input scale of \cite{grv92} as a lower and the physical
scale $Q^2$ as an upper limit.}
\end{figure}  
In Fig.\ \ref{switch2} we cover the maximally conceivable range  
for $Q_0^2$; i.e.\ the leftmost end of all curves is set by the
low input scale of the parton distributions in \cite{grv92},   
$Q_0^2=0.3\ {\rm{GeV}}^2$, while the rightmost ends are 
the `fixed order limit' in Eq.\ (\ref{foptlim}). 
The observed monotonic scale dependence has to be expected from the 
positivity of the collinear resummation at small $x$ which is
continuously suppressed the more $Q_0$ is increased. Worrisome is, however, the
steep slope $\partial F_2^{c,MP}/ \partial \ln (Q_0^2/m_c^2)$
around $Q_0\sim m_c$ 
for high $W^2=Q^2 (1/x-1)$, 
where the charm contribution is most
important. This observation restricts the predictive power of NLO 
collinear resummation techniques
which have thus far been constructed to match
the asymptotic ($Q^2\rightarrow \infty$) $F_2^{c,MP}$ derived from 
$c(x,Q_0^2=m_c^2)=0$. 
The uncertainty from the residual $Q_0$ dependence,
inherent to any VFNS \cite{thoro,acot,mrrs} worked out to NLO,
seems to dominate over scheme uncertainties 
of ${\cal O}(m_c^2/Q^2)$ between the individual schemes.
The arbitrariness of $Q_0$ 
therefore constitutes 
a main limiting factor on the perturbative accuracy of VFNS 
heavy quark predictions at high $Q^2$. 
On the other hand we observe a flattening slope 
$\partial F_2^{c,MP}/ \partial \ln (Q_0^2/m_c^2)$
towards the `fixed order limit' at high $Q_0 \lesssim Q$ where perturbative 
NLO $\leftrightarrow$ LO stability had been found in \cite{grs}
by $K$ factor considerations for the {\it full} fixed order predictions, 
i.e.\ including logs {\it and} finite terms. We should reemphasize that
these conclusions are based on the NLO matching conditions in Eq.\ (\ref{bound})
- which is the present state of the art for PDF sets including partonic 
heavy quarks \cite{grv92,ref:mrst,cteq5} - 
and do not take into account
the higher corrections of \cite{bmsn,alpsnnlo}. 
The terms beyond Eq.\ (\ref{bound}) 
represent NNLO contributions 
to the asymptotic VFNS prediction $F_2^{c,MP}$. 
Very recently the results of \cite{bmsn} have been implemented in a 
\alpsq\ implementation of a VFNS\footnote{A $c(x,Q^2)$ derived from the NNLO boundary 
conditions in \cite{bmsn} would, however, be problematic to apply to hadroproduction 
calculations, since the higher terms in \cite{bmsn} are not yet contained 
in, e.g., the fixed NLO [${\cal{O}}(\alpha_s^3)$] hadroproduction process
$p{\bar p}\rightarrow c(p_T) X$ \cite{hadroprod}.}  
where the contribution from the unknown NNLO (3-loop) splitting functions 
had to be neglected. Choosing $Q_0=m$ the results of \cite{smith}
seem to indicate that the impact of terms from the resummation beyond fixed NLO 
[\alpsq ]\footnote{The confusion
of counting perturbative orders differently in resummed (MP) and fixed order
(FOPT) calculations is treated in more detail in \cite{thoro,collins}.}
perturbation theory is rather moderate.
As a further step in the line of the present investigation 
it would clearly be interesting to generalize \cite{smith}
to $Q_0\neq m$. If a $Q_0>m$ would be prefered by such an analysis the difference 
between VFNS and FOPT would be reduced even more.
Such a result would again re-confirm the perturbative reliability of FOPT found
in \cite{grs} as much as it would help reduce unphysical scheme
dependences of QCD predictions on charm production and thus make a comparison
to experiment even more compelling.    

To summarize, we have considered variations of the {\it a priori} arbitrary   
charm input scale $Q_0$, 
which separates a 3 from a 4 flavour scheme in variable flavour number 
approaches, 
around its usually adopted but by no means theoretically required value of
$Q_0=m_c$. From the NLO boundary conditions 
we found a monotonic $Q_0$ dependence and a worrisome 
steep slope of 
$F_2^{c,MP}$ in (\ref{mp}) with respect to $\ln Q_0^2$ just around 
$Q_0 \sim m_c$. 
This behaviour restricts the accuracy of a collinearly resummed
NLO approach towards calculating high $Q^2$ charm production   
from matching \oalps\ boson gluon fusion at $Q_0=m_c$
in the MP component of Eq.~(\ref{simple}).
This uncertainty in MP from the unknown $Q_0$ feeds back onto the entire VFNS
via Eq.\ (\ref{simple}) which is normalized to MP asymptotically
($Q^2\gg m^2$). 
Our results imply to consider     
variations of $Q_0$ both as a limiting factor on the present perturbative accuracy if 
estimating the theoretical uncertainty of VFNS heavy quark predictions
as well as in order to optimize the starting scale for the charm evolution
within 
higher order realizations of VFNSs \cite{bmsn,smith,cschm}.
This latter higher order analysis assumes, however, that the unknown NNLO (3-loop)
splitting functions can be neglected \cite{smith}.

\section*{Acknowledgements}
We thank E.\ Reya and I.\ Schienbein for valuable discussions and for
carefully reading the manuscript.
The work has been supported in part by the 
`Bundesministerium f\"{u}r Bildung, Wissenschaft, Forschung und
Technologie', Bonn. 

%


\end{document}